\documentclass[amsmath,amssymb,superscriptaddress,nobalancelastpage,aps,prl,twocolumn]{revtex4-1}

\usepackage{graphicx}
\usepackage{varioref}
\usepackage{xr-hyper}
\usepackage{xcolor}
\usepackage{nicefrac}
\usepackage{xfrac}
\usepackage{hyperref}
\hypersetup{colorlinks,linkcolor=blue,urlcolor=blue,citecolor=blue}
\usepackage{ulem}
\usepackage{lineno}
\usepackage{amsmath}
\usepackage{amssymb}
\usepackage{bm}

\newcommand{\EuLSCO}{La$_{1.675}$Eu$_{0.2}$Sr$_{0.125}$CuO$_4$}
\newcommand{\NdLSCO}{La$_{1.475}$Nd$_{0.4}$Sr$_{0.125}$CuO$_4$}
\newcommand{\LBCO}{La$_{1.875}$Ba$_{0.125}$CuO$_4$}
\newcommand{\LSCO}{La$_{1.875}$Sr$_{0.125}$CuO$_4$}
\begin{document}

\title{High-Temperature Charge-Stripe Correlations in La$_{1.675}$Eu$_{0.2}$Sr$_{0.125}$CuO$_4$}

\author{Qisi~Wang}
\email{qisiwang@physik.uzh.ch}
\affiliation{Physik-Institut, Universit\"{a}t Z\"{u}rich, Winterthurerstrasse
190, CH-8057 Z\"{u}rich, Switzerland}

\author{M.~Horio}
\affiliation{Physik-Institut, Universit\"{a}t Z\"{u}rich, Winterthurerstrasse
190, CH-8057 Z\"{u}rich, Switzerland}

\author{K.~von~Arx}
\affiliation{Physik-Institut, Universit\"{a}t Z\"{u}rich, Winterthurerstrasse
190, CH-8057 Z\"{u}rich, Switzerland}

\author{Y.~Shen}
\affiliation{State Key Laboratory of Surface Physics and Department of Physics, Fudan University, Shanghai 200433, China}

\author{D.~John~Mukkattukavil}
\affiliation{Department of Physics and Astronomy, Uppsala University, Box 516, SE-751 20 Uppsala, Sweden}

\author{Y.~Sassa}
\affiliation{Department of Physics, Chalmers University of Technology, SE-412 96 G\"{o}teborg, Sweden}

\author{O.~Ivashko}
\affiliation{Physik-Institut, Universit\"{a}t Z\"{u}rich, Winterthurerstrasse 190, CH-8057 Z\"{u}rich, Switzerland}

\author{C.~E.~Matt}
\affiliation{Swiss Light Source, Photon Science Division, Paul Scherrer Institut, CH-5232 Villigen PSI, Switzerland}
\affiliation{Physik-Institut, Universit\"{a}t Z\"{u}rich, Winterthurerstrasse 190, CH-8057 Z\"{u}rich, Switzerland}

\author{S.~Pyon}
\affiliation{Department of Advanced Materials, University of Tokyo, Kashiwa 277-8561, Japan}

\author{T.~Takayama}
\affiliation{Department of Advanced Materials, University of Tokyo, Kashiwa 277-8561, Japan}

\author{H.~Takagi}
\affiliation{Department of Advanced Materials, University of Tokyo, Kashiwa 277-8561, Japan}

\author{T.~Kurosawa}
\affiliation{Department of Physics, Hokkaido University, Sapporo 060-0810, Japan}

\author{N.~Momono}
\affiliation{Department of Physics, Hokkaido University, Sapporo 060-0810, Japan}
\affiliation{Department of Applied Sciences, Muroran Institute of Technology, Muroran 050-8585, Japan}

\author{M.~Oda}
\affiliation{Department of Physics, Hokkaido University, Sapporo 060-0810, Japan}

\author{T.~Adachi}
\affiliation{Department of Engineering and Applied Sciences, Sophia University, Tokyo 102-8554, Japan}

\author{S.~M.~Haidar}
\affiliation{Department of Applied Physics, Tohoku University, Sendai 980-8579, Japan}

\author{Y.~Koike}
\affiliation{Department of Applied Physics, Tohoku University, Sendai 980-8579, Japan}

\author{Y.~Tseng}
\affiliation{Swiss Light Source, Photon Science Division, Paul Scherrer Institut, CH-5232 Villigen PSI, Switzerland}

\author{W.~Zhang}
\affiliation{Swiss Light Source, Photon Science Division, Paul Scherrer Institut, CH-5232 Villigen PSI, Switzerland}

\author{J.~Zhao}
\affiliation{State Key Laboratory of Surface Physics and Department of Physics, Fudan University, Shanghai 200433, China}
\affiliation{Collaborative Innovation Center of Advanced Microstructures, Nanjing 210093, China}

\author{K.~Kummer}
\affiliation{European Synchrotron Radiation Facility, 71 Avenue des Martyrs, 38043 Grenoble, France}

\author{M.~Garcia-Fernandez}
\affiliation{Diamond Light Source, Harwell Campus, Didcot, Oxfordshire OX11 0DE, United Kingdom}

\author{Ke-Jin~Zhou}
\affiliation{Diamond Light Source, Harwell Campus, Didcot, Oxfordshire OX11 0DE, United Kingdom}

\author{N.~B.~Christensen}
\affiliation{Department of Physics, Technical University of Denmark, DK-2800 Kongens Lyngby, Denmark}

\author{H.~M.~R{\o}nnow}
\affiliation{Institute of Physics, \'{E}cole Polytechnique Fed\'{e}rale de Lausanne (EPFL), CH-1015 Lausanne, Switzerland}

\author{T.~Schmitt}
\affiliation{Swiss Light Source, Photon Science Division, Paul Scherrer Institut, CH-5232 Villigen PSI, Switzerland}

\author{J.~Chang}
\email{johan.chang@physik.uzh.ch}
\affiliation{Physik-Institut, Universit\"{a}t Z\"{u}rich, Winterthurerstrasse 190, CH-8057 Z\"{u}rich, Switzerland}

\begin{abstract}
We use resonant inelastic x-ray scattering to investigate charge-stripe correlations in La$_{1.675}$Eu$_{0.2}$Sr$_{0.125}$CuO$_4$. By differentiating elastic from inelastic scattering, it is demonstrated that charge-stripe correlations precede both the structural low-temperature tetragonal phase and the transport-defined pseudogap onset. The scattering peak amplitude from charge stripes decays approximately as $T^{-2}$ towards our detection limit. The in-plane integrated intensity, however, remains roughly temperature independent.
Therefore, although the incommensurability shows a remarkably large increase at high temperature, our results are interpreted via a single scattering constituent. In fact, direct comparison to other stripe-ordered compounds (La$_{1.875}$Ba$_{0.125}$CuO$_4$, La$_{1.475}$Nd$_{0.4}$Sr$_{0.125}$CuO$_4$ and La$_{1.875}$Sr$_{0.125}$CuO$_4$) suggests a roughly constant integrated scattering intensity across all these compounds. Our results therefore provide a unifying picture for the charge-stripe ordering in La-based cuprates. As charge correlations in La$_{1.675}$Eu$_{0.2}$Sr$_{0.125}$CuO$_4$ extend beyond the low-temperature tetragonal and pseudogap phase, their emergence heralds a spontaneous symmetry breaking in this compound.
\end{abstract}

\maketitle

\begin{figure*}[ht!]
 	\begin{center}
 		\includegraphics[width=0.95\textwidth]{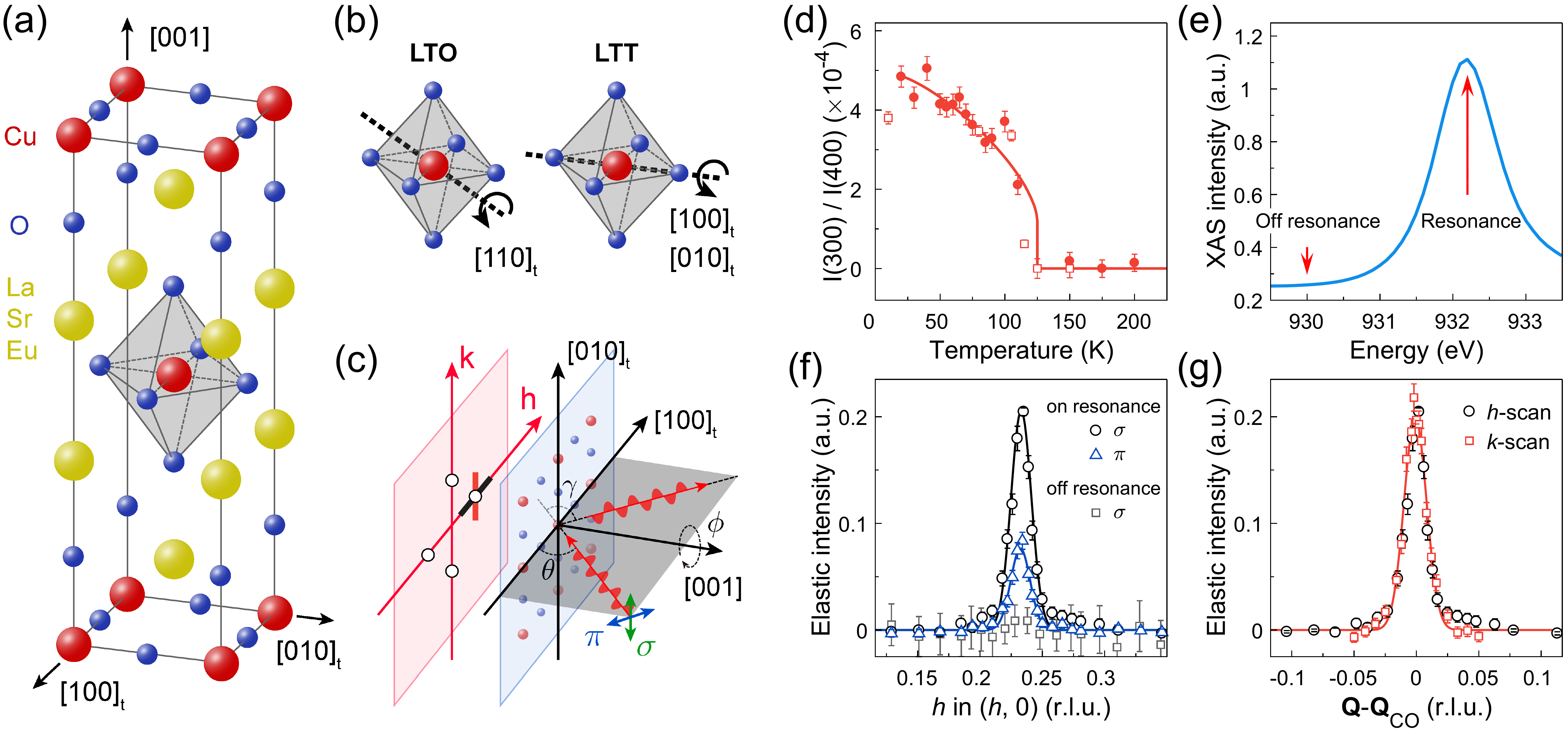}
 	\end{center}
 	\caption{Crystal structure and low-temperature charge order in LESCO. (a) Crystal structure of LESCO in the high-temperature tetragonal phase. Upon cooling, it first undergoes a structural transition to the low-temperature orthorhombic (LTO) phase then enters the LTT phase at lower temperature. (b) Distortions of the CuO$_{6}$ octahedra in the LTO and LTT phases. (c) Schematic of the scattering geometry of the RIXS experiments. The incident angle $\theta$ is defined with respect to the CuO$_{2}$ plane. The black open circles denote the in-plane position of the CO reflections. (d) Temperature dependence of $(3,0,0)$ structural peak intensity, measured with nonresonant 8.048~keV x rays, characterizes the development of LTT phase. The intensity is normalized to the amplitude of $(4,0,0)$ Bragg peak. The filled and open symbols represent data measured by counting at both the peak and background positions or by fitting longitudinal scans, respectively. (e) Normal incidence x-ray absorption spectroscopy (XAS) spectrum obtained with $\sigma$ polarized light. (f) Momentum scans ($\mathbf{Q}$-scans) of the elastic intensity through the CO peak position along $h$ measured with $\sigma$ and $\pi$ polarizations and at the off-resonance energy (930.0~eV) indicated in (e). (g) Elastic $\mathbf{Q}$-scans along $h$ and $k$ directions as indicated by the black and red thick lines in (c), respectively. Solid lines in (f) and (g) are Gaussian fits. Error bars are defined by counting statistics or estimated from systematic uncertainties throughout the Letter~[\hyperref[fn:fn1]{20}].
 	}	
	 	\label{fig:fig1}
 \end{figure*}

Unconventional superconductivity is often associated with competing intertwined order parameters. For underdoped cuprate superconductors, the omnipresence of charge ordering (CO) has been established~\cite{tranquada,ZhangPRL2018,chang12,Wu11,GhiringhelliSCI12}. The observation of different ordering vectors, however, opens the possibility of multiple ordering susceptibilities. It is therefore pivotal to investigate the different types of charge ordering tendencies.
In this context, the La-based cuprates constitute an important group of compounds. In fact, spin-charge stripe order was first discovered in \NdLSCO~(LNSCO)~\cite{tranquada} and subsequently studied intensively in the La$_{2-x}$Ba$_{x}$CuO$_4$ system~\cite{FujitaPRB2004,hucker,HuckerPRB2011}. These two compounds share a low-temperature tetragonal (LTT) crystal structure that appears concomitantly with charge ordering. The LTT phase is characterized by tilting of the CuO$_{6}$ octahedra with the rotation axis along the copper-oxygen bond directions [see Figs.~\ref{fig:fig1}(a) and \ref{fig:fig1}(b)]. It breaks $C_{4}$ symmetry within individual CuO$_{2}$ planes, consistently with charge-stripe ordering. The potential link between charge-stripe ordering and the LTT phase has therefore been the subject of many experiments~\cite{tranquada,HuckerPRB2011,HuckerPRL2010,ZhangPRL2018,AchkarSCI2016}. A fundamental question is whether the LTT structure is a consequence or trigger of stripe order.
The original scattering experiments found that the LTT and stripe order parameters have identical onsets in temperature~\cite{tranquada,FujitaPRB2004}. As the LTT order parameter grows faster, stripe order has been viewed as a secondary effect.
Starting from \LBCO~(LBCO), another set of experiments has traced stripe ordering as the LTT phase is suppressed by hydrostatic pressure~\cite{HuckerPRL2010}, temperature~\cite{MiaoPNAS2017}, or Sr for Ba substitution~\cite{ThampyPRB2014,CroftPRB2014,Christensen14}. In all cases, it is found that charge-stripe order emerges spontaneously outside the LTT crystal lattice phase. A prominent exception is \EuLSCO~(LESCO) that has the highest LTT onset temperature [$T_{\rm{s}} \sim 125$~K; see Fig.~\ref{fig:fig1}(d)] and where stripe order, probed by resonant elastic x-ray scattering (REXS), appears only below 80~K~\cite{FinkPRB2009,FinkPRB2011,AchkarSCI2016}.
This pronounced temperature ``gap" between the LTT and REXS-defined stripe order onsets suggests no fundamental correlation between the two. The lack of universal behavior across the La-based cuprates has impeded broader conclusions on the stripe ordering.

Here, we present a high-resolution resonant inelastic x-ray scattering (RIXS) study of La$_{1.675}$Eu$_{0.2}$Sr$_{0.125}$CuO$_{4}$ (LESCO-1/8).
By filtering inelastic from elastic processes, it is shown that charge correlations survive beyond both the LTT phase and the pseudogap onset $T$* defined by transport experiments~\cite{PGPRB2018}.
This demonstrates that, in this compound, stripe correlations appear spontaneously in the absence of LTT and pseudogap phase.
The charge-stripe diffraction peak amplitude decays approximately as $T^{-2}$ as our RIXS detection limit is approached in the high-temperature limit. The diffraction intensity integrated within the probed momentum space, by contrast, appears roughly temperature independent. Although the charge ordering wave vector exhibits a large shift at high temperature, the constant integrated intensity suggests a single ordering mechanism. Finally, we provide evidence for an approximately constant in-plane integrated intensity across all known 1/8-doped stripe-ordered cuprate systems.

 \begin{figure*}[ht!]
 	\begin{center}
 		\includegraphics[width=0.9\textwidth]{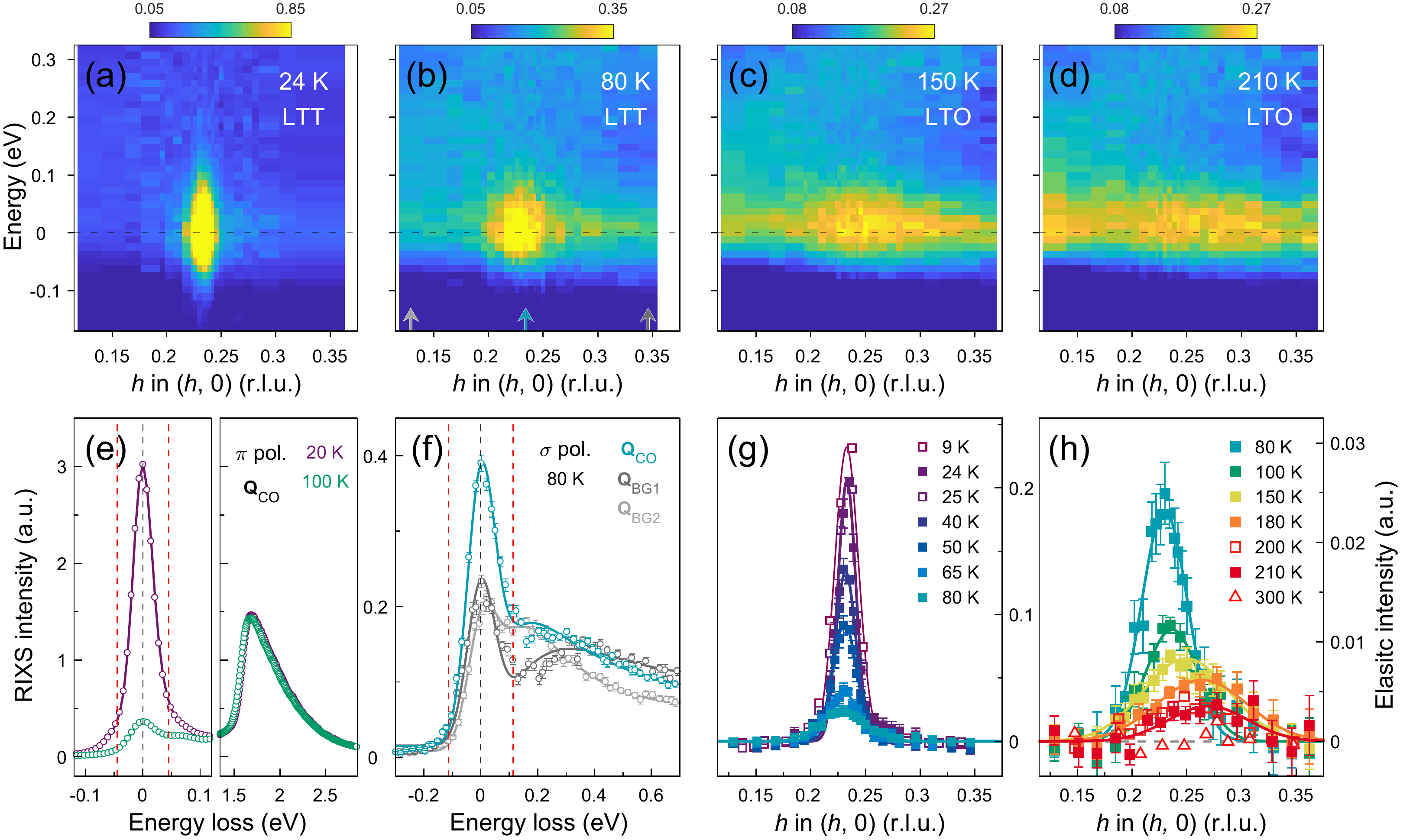}
 	\end{center}
 	\caption{Temperature evolution of the charge-stripe order. (a)--(d) Intensity distribution maps of the raw RIXS spectra in the energy-momentum space through the CO peak for different temperatures. (e) High-resolution RIXS spectra obtained near $\mathbf{Q}\rm_{CO}$ with $\pi$ polarization for indicated temperatures. Data near the elastic line (left-hand part) and $dd$ excitations (right-hand part) are shown on different energy scales for clarity. (f) Representative RIXS spectra collected at 80~K. The blue, black and light gray open circles represent data measured at the CO peak and background positions indicated by the arrows in (b) [$\mathbf{Q}\rm_{BG1}=(0.346,0)$, $\mathbf{Q}\rm_{BG2}=(0.129,0)$], respectively. The red dashed lines in (e) and (f) mark the energy window of the elastic intensity integration. (g), (h) Elastic scans along $h$ at various temperatures with linear backgrounds subtracted. The filled and open symbols represent data obtained at ADRESS and I21, respectively. Solid lines in (g) and (h) are Gaussian fits to the data.
 	}
	 	\label{fig:fig2}
 \end{figure*}

Single crystals of LESCO, LNSCO, LBCO, and \LSCO~(LSCO) were grown using a floating zone method. RIXS experiments were carried out at the ADRESS~\cite{ghiringhelliREVSCIINS2006,strocovJSYNRAD2010}, I21, and ID32 beam lines at the Swiss Light Source (SLS) at the Paul Scherrer Institut, Diamond Light Source, and European Synchrotron Radiation Facility (ESRF), respectively.
Energy resolution, expressed in standard Gaussian deviation ranges from $\sigma_{_G}\approx19$ to $48$~meV for experiments on LESCO. A complete discussion of experimental configurations and resolution linewidth profiles is given in the Supplemental Material~\footnote{\label{fn:fn1}See Supplemental Material for details on experimental configurations, fitting of RIXS spectra, data normalization, background subtraction, temperature dependence of the inelastic spectral weight and characterization of the LTT structural transition, which includes Refs.~\cite{AmentRMP2011,AmentPRL09,HaverkortPRL10,SalaNJP,ZimmermannEPL98,KimuraPRB2003,MiaoPNAS2017,Kim09,ThampyPRB2014,WilkinsPRB2011,HuckerPRB2011,SimovicPRB2003}}. Given the quasi-two-dimensional character of this system, only the in-plane momentum is considered and varied through the incident light $\theta$ and sample azimuthal $\phi$ angles~[Fig.~\ref{fig:fig1}(c)].
Both linear vertical ($\sigma$) and linear horizontal ($\pi$) incident light polarizations [Fig.~\ref{fig:fig1}(c)] were used.
All crystals were cleaved \textit{in-situ} by a standard top-post technique.
Wave vector $\mathbf{Q}$ at $(q_{x},q_{y},q_{z})$ is defined as $(h,k,\ell)=(q_{x}a/2\pi,q_{y}b/2\pi,q_{z}c/2\pi)$ reciprocal lattice units (r.l.u.) using pseudo-tetragonal notation, with $a\approx b\approx3.79$~\AA~and $c\approx13.1$~\AA ~for LESCO.

Charge-stripe order in the La-based cuprates manifests itself by satellite peaks at the wave vectors $\mathbf{Q}=\bm{\tau}+\mathbf{Q}\rm_{CO}$ where $\bm{\tau}$ represents fundamental Bragg reflections and $\mathbf{Q}\rm_{CO}=(\delta_a, 0, \delta_c)$ and $(0,\delta_b,\delta_c)$. The in-plane incommensurability $\delta=\delta_a=\delta_b$ is close to 1/4, and weak out-of-plane anti-phase correlations imply broad diffraction maxima at $\delta_c=1/2$~\cite{HuckerPRB2011,Christensen14}.
Cu $L$-edge [$\sim932.2$~eV, Fig.~\ref{fig:fig1}(e)] RIXS permits resonant x-ray diffraction (XRD) with a well-defined energy resolution.
Such resonant energy-resolved XRD data recorded on LESCO-1/8 at $T=24$~K are shown in Figs.~\ref{fig:fig1}(f) and \ref{fig:fig1}(g). It reveals -- consistent with previous reports~\cite{FinkPRB2009,FinkPRB2011,AchkarSCI2016} -- that (i) the low-temperature charge order incommensurability is $\delta=0.233$~\cite{FinkPRB2009,Peng19}, (ii) the electronic ordering is probed on the resonance only~\cite{MiaoPNAS2017}, (iii) the reflection is most intense using incident $\sigma$ polarization~\cite{GhiringhelliSCI12}, and (iv) the in-plane transverse and longitudinal correlation lengths are identical~\cite{MiaoPNAS2017,WilkinsPRB2011}. Unless otherwise indicated, we therefore use $\sigma$ polarization and focus on in-plane longitudinal $(h,0)$ scans.

Temperature dependence of the charge scattering is shown in Figs.~\ref{fig:fig2}(a)--\ref{fig:fig2}(d), by plotting RIXS intensity distribution maps (IDMs) -- intensity versus $h$ and energy loss. The main experimental observation reported here is that the charge-stripe correlations in LESCO persist beyond the temperature onset reported by REXS~\cite{FinkPRB2009,FinkPRB2011,AchkarSCI2016} and even beyond the onset temperature of the LTT phase [Fig.~\ref{fig:fig1}(d)]. This is directly demonstrated by the raw RIXS IDMs as a function of temperature.
Representative RIXS spectra measured with both the $\pi$ and $\sigma$ polarizations are shown in Figs.~\ref{fig:fig2}(e) and \ref{fig:fig2}(f), respectively. The elastic charge scattering is drastically enhanced when cooling to base temperature [Fig.~\ref{fig:fig2}(e)] and by approaching $\mathbf{Q}\rm_{CO}$~[Fig.~\ref{fig:fig2}(f)].

 \begin{figure*}[ht!]
 \begin{minipage}{\textwidth}
 	\begin{center}
 		\includegraphics[width=0.9\textwidth]{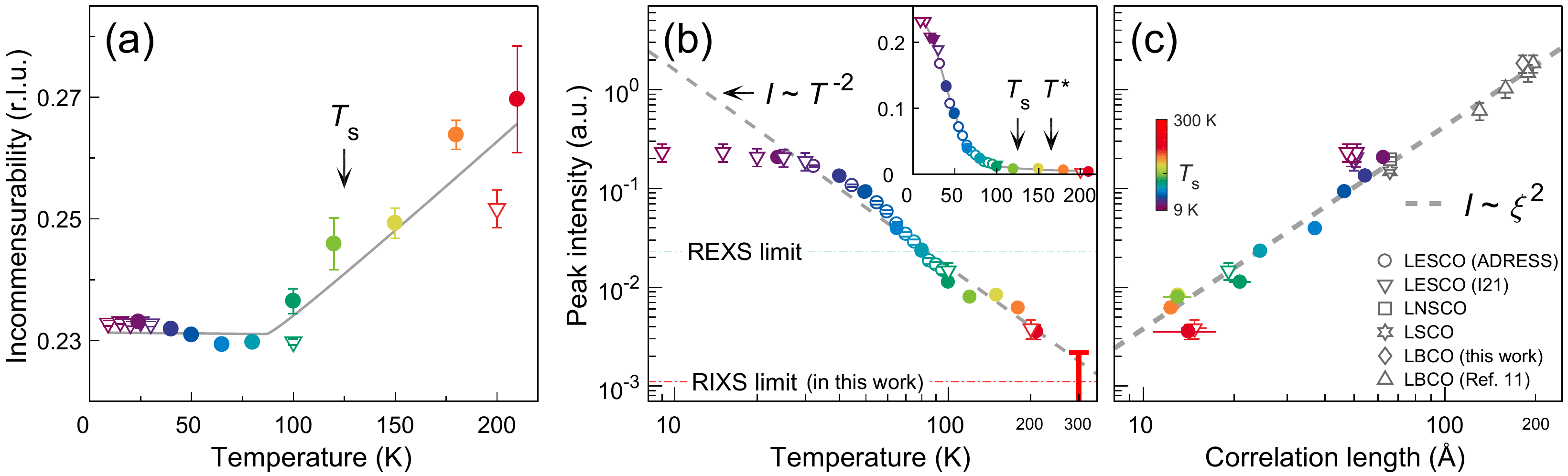}
 	\end{center}
 	\caption{Incommensurability, intensity and correlation length of the charge-stripe order. (a), (b) Temperature dependence of the CO incommensurability (a) and peak intensity (b) in LESCO. The filled circles and open triangles in (b) denote the fitted amplitude of $h$-scans obtained at ADRESS and I21, respectively. The open circles represent the background-subtracted elastic scattering intensity measured at $\mathbf{Q}=(0.235,0)$. Backgrounds are measured at $\mathbf{Q}=(0.129,0)$ and $(0.346,0)$. The red vertical bar at 300 K illustrates the upper limit of the charge order peak height estimated from the $h$-scan. The inset shows the same data on a linear scale. The REXS~\cite{FinkPRB2009,FinkPRB2011,AchkarSCI2016} and RIXS detection limits are indicated by horizontal dash-dotted lines. (c) CO peak intensity as a function of correlation length for LESCO, LNSCO, LBCO and LSCO. Since the CO reflection in LSCO splits transversely into two peaks at low temperature (see Fig.~S2(d)~[\hyperref[fn:fn1]{20}] and Ref.~\cite{ThampyPRB2014}), twice of the amplitude of a single peak is used for LSCO. The color-code indicates the temperature for the LESCO data. Solid lines are guides to the eye.
 	}	
	 	\label{fig:fig3}
   \end{minipage}
 \end{figure*}

To quantify the charge order intensity, these RIXS IDMs are analyzed as follows. Elastic scattering is inferred from each RIXS spectrum by integrating the intensity $\pm2\sqrt{2\ln{2}}\sigma_{_G}$ around the elastic line [see Fig.~\ref{fig:fig2}(e)].
To account for the variation of detection efficiency, the RIXS intensities are normalized to the weight of the $dd$ excitations. In this fashion, after subtracting a linear background~\footnotemark[\value{footnote}], longitudinal scans through $\mathbf{Q}\rm_{CO}$ [see Figs.~\ref{fig:fig1}(f), \ref{fig:fig2}(g) and \ref{fig:fig2}(h)] are generated. Gaussian fits of these scans allow us to extract (i) the scattering amplitude $I$, (ii) the incommensurability $\delta$ and (iii) the in-plane correlation length $\xi$ defined by the inverse half width at half maximum (HWHM). The temperature evolution of these parameters are summarized in Fig.~\ref{fig:fig3}.

The incommensurability is locked to $\delta\approx 0.23$ below 80~K. By contrast for $T>80$~K, it increases upon warming and reaches $\delta\approx 0.27$ at $T=210$~K [Fig.~\ref{fig:fig3}(a)].
The diffraction amplitude $I$ decreases initially ($T<30$~K) slowly before an approximately power-law $T^{-2}$ decay [Fig.~\ref{fig:fig3}(b)].
As $\xi$ decays in a similar fashion with temperature, plotting $I$ versus $\xi$ reveals a simple power-law relation between the two \textit{i.e.}, $I\sim \xi^\alpha$ with $\alpha\approx2.0$ [Fig.~\ref{fig:fig3}(c)].
The similar crystal field environment across La-based single-layer cuprates justifies comparison of CO diffraction intensities normalized to their respective $dd$ excitations. Doing so, the $I\sim \xi^2$ scaling applies also to LNSCO, LBCO and LSCO with doping $x=1/8$ [see Fig.~\ref{fig:fig3}(c)].

We start by discussing how our RIXS experiments on LESCO compare to previous REXS reports~\cite{FinkPRB2011,FinkPRB2009,AchkarSCI2016}.
The difference in charge ordering ``onset" temperature found by, respectively, REXS and RIXS is a matter of probing sensitivity.
With the energy resolving power of RIXS, it is possible to disentangle elastic from inelastic scattering processes~\cite{AmentRMP2011}.
As the diffraction amplitude from charge order is weakened upon increasing temperature, it eventually becomes negligible in comparison to, for example, $dd$ excitations [Fig.~\ref{fig:fig2}(e)]~\cite{SalaNJP}.
In LESCO, this happens around $T\approx80$~K, and hence this temperature scale marks the charge correlation detection limit for REXS experiments rather than a fundamental onset temperature. With differentiation of elastic and inelastic scattering processes, it is possible to probe the charge ordering response in LESCO in the regime beyond the REXS detection limit.
The CO diffraction amplitude decays roughly as $T^{-2}$ until the RIXS detection limit is reached between 200 and 300 K.
Even more intriguingly, the in-plane integrated diffraction intensity remains essentially independent of temperature \textit{i.e.}, $I\propto \xi^2$.
Cooling into the pseudogap and LTT phases thus has no significant impact on the stripe order intensity. It is thus clear that the first order onset of LTT structure is neither triggering nor enhancing the stripe order in LESCO-1/8.

The scaling behavior of charge scattering ($I\propto \xi^2$) is reminiscent of the dynamic magnetic critical scattering in La$_2$CuO$_4$ and another $S=1/2$ two-dimensional square-lattice quantum Heisenberg antiferromagnet (2DSLQHA) Sr$_2$CuO$_2$Cl$_2$ observed by neutron scattering experiments~\cite{BirgeneauPRB99,GrevenZPB95}.
The amplitude of instantaneous spin correlations $S(\mathbf{q})=S(0)/[1+(\mathbf{q}/\xi)^2]$ is also found to scale with the spin correlation length as $S(0)\propto \xi^2$. Here $\mathbf{q}$ is the two-dimensional deviation from the antiferromagnetic ordering wave vector. Notably, such scaling behavior observed on La$_2$CuO$_4$ is associated with two-dimensionality and local spin nature.
However, it seems to be specific to the cuprates, as in copper formate tetradeuterate (CFTD), another $S=1/2$ 2DSLQHA, the ratio of $S(0)/\xi^2$ shows a clear temperature dependence~\cite{RnnowPRL99}.
As the response function for charge correlations is represented in the low-energy RIXS cross section~\cite{AmentRMP2011,JiaPRX16}, our experiments on stripe-ordered cuprates are analogous to the neutron scattering studies of La$_2$CuO$_4$. Drawing on this analog may suggest that critical charge correlations are probed.

The incommensurability $\delta$ in LESCO-1/8 is displaying, as in LBCO-1/8~\cite{MiaoPNAS2017,MiaoPRX2019}, a strong temperature dependence.
Below 80~K, a low-temperature lock-in of the incommensurability occurs. This lock-in temperature scale seems unrelated to the LTT structural transition.
The fact that the incommensurability $\delta$ moves -- with increasing temperature ($T>80$~K) -- from $\approx 1/4$ toward the $\approx 1/3$ found in YBa$_2$Cu$_3$O$_{7-y}$ (YBCO)~\cite{Wu11,GhiringhelliSCI12,chang12,AchkarPRL2012,HuckerPRB2014}, led to the suggestion that there exists a universal susceptibility for charge order correlations with $\delta\sim 0.3$ in hole-doped cuprates~\cite{MiaoPNAS2017}. However, in LESCO the charge order incommensurability is not reaching $\delta=0.3$ and hence the link to charge ordering in YBCO remains speculative. Future ultrahigh-resolution RIXS experiments should address whether the apparent shift in incommensurability is influenced by soft phonons or Kohn anomalies near $\bf{Q}\rm_{CO}$~\cite{BlackburnPRB2013,TaconNatPhys2014,ReznikNat2006}. If such phonon excitations become relevant at high temperature, one would expect a broadening of the elastic CO peak. This is indeed observed, but there seems to be no obvious correlation between the broadening and incommensurability.

In both YBCO~\cite{Arpaia906} and LBCO~\cite{MiaoPRX2019}, the temperature dependence of the charge ordering has been analyzed using two components: a static long-range low-temperature and a fluctuating short-range high-temperature constituent. Our data on LESCO are analyzed with a single component that has a constant in-plane integrated intensity. Despite different analysis approaches, the low-temperature long-range correlations and integrated diffraction intensity can certainly be compared across 1/8-doped stripe-ordered compounds. The longest in-plane correlation lengths are found in LBCO~\cite{MiaoPNAS2017,WilkinsPRB2011}, whereas more modest length scales are reported in LNSCO~\cite{tranquada,WilkinsPRB2011,ZimmermannEPL98,Kim09}, LSCO~\cite{ThampyPRB2014} and LESCO (this work). Normalizing, the best possible, intensities across all these four compounds suggests a roughly constant in-plane integrated intensity (see Fig.~S3 in Supplemental Material \footnotemark[\value{footnote}])~\footnote{\label{fn:fn2}The previous REXS work in Ref.~\cite{WilkinsPRB2011} suggests that the in-plane integrated charge ordering intensity in LBCO is $\sim$~10 times larger than that in LNSCO. However, we note that compared to the current work and other studies~\cite{ZimmermannEPL98,Kim09}, a much larger in-plane correlation length ($\xi = 111\pm7$~\AA) of charge order in LNSCO is reported in Ref.~\cite{WilkinsPRB2011}.}. The fact that the LTT onset temperature and order parameter vary  across these compounds~\cite{AchkarSCI2016,FinkPRB2011,WilkinsPRB2011} again suggests no link between the diffracted stripe order and  the structural phase transition. It is also interesting to discuss correlation length and integrated diffracted intensity in the context of the Fermi surface reconstruction (FSR). In LBCO, LNSCO and LESCO, the large low-temperature negative thermopower has been interpreted as evidence of a charge-stripe-order induced FSR~\cite{ChangPRL2010,LaliberteNatComm2011}. This is supported by the fact that the negative thermopower is found in a dome around 1/8 doping~\cite{LaliberteNatComm2011}. Typically, the onset temperature $T\rm_{FSR}$ of the FSR is quantified by the zero crossing ($S=0$) of the thermopower~\cite{LaliberteNatComm2011}. Both $T\rm_{FSR}$ and the zero-temperature thermopower values are essentially identical across these stripe-ordered compounds. The identical FSR onset temperature suggests that the in-plane correlation length is not the defining parameter for the FSR. Empirically, the in-plane integrated scattering intensity seems to be a more important parameter.

In summary, we have carried out resonant inelastic x-ray scattering experiments on stripe-ordered La-based cuprates. Focusing on LESCO-1/8, it is demonstrated -- in contrast to previous elastic (resonant and nonresonant) reports -- that stripe correlations precede the low-temperature tetragonal crystal structure. This permits us to conclude that stripe order generally emerges spontaneously without LTT structural deformations. While charge correlations in LESCO have been traced to the highest temperature among the stripe-ordered cuprates, the integrated scattering intensities are found to be compound independent. This, together with the temperature dependence of the diffraction amplitude, strongly suggests that stripe ordering in the cuprates goes beyond weak-coupling mean-field physics.

\begin{acknowledgments}
We thank T.~M.~Rice, L.~Taillefer, M.-H.~Julien and M.~H\"{u}cker for insightful discussion and Z.~He and Y.~Hao for assistance with XRD measurements. Q.W., M.H., K.v.A., C.E.M., Y.T., W.Z., T.S., and J.C. acknowledge support by the Swiss National Science Foundation through the SINERGIA network Mott Physics Beyond the Heisenberg Model and Grants No. BSSGI0-155873, No. P400P2\_183890, and No. 200021-178867.
Y.~Sassa is funded by the Swedish Research Council (VR) with a Starting Grant (Dnr. 2017-05078).
Y.~Shen and J.Z. were supported by the Innovation Program of Shanghai Municipal Education Commission (Grant No. 2017-01-07-00-07-E00018) and the National Key R\&D Program of the MOST of China (Grant No. 2016YFA0300203).
T.A. was supported by JSPS KAKENHI through Grant No. JP19H01841.
N.B.C. thanks the Danish Agency for Science, Technology, and Innovation for funding the instrument center DanScatt.
We acknowledge Swiss Light Source at the Paul Scherrer Institut, Diamond Light Source, and European Synchrotron Radiation Facility for approval of RIXS proposals at the ADRESS (No.~20182238), I21 (No.~SP20828, MM24481), and ID32 (No.~HC3888) beam lines, respectively.

\end{acknowledgments}

\end{document}